\documentclass[twocolumn,floatfix,superscriptaddress,amsmath,showpacs,aps,pre]{revtex4-1}
\begin{document}
\title{Supersymmetric formulation of  multiplicative white--noise stochastic processes}
\author{Zochil Gonz\'alez Arenas}
\affiliation{Instituto de Cibern\'etica, Matem\'atica y F\'\i sica (ICIMAF)\\
Calle 15 \# 551  e/ C y D, Vedado, C. Habana, Cuba.}
\affiliation{Departamento de F{\'\i}sica Te\'orica,
Universidade do Estado do Rio de Janeiro, Rua S\~ao Francisco Xavier 524, 20550-013,  Rio de Janeiro, RJ, Brazil.}
\author{Daniel G.\ Barci}
\affiliation{Departamento de F{\'\i}sica Te\'orica,
Universidade do Estado do Rio de Janeiro, Rua S\~ao Francisco Xavier 524, 20550-013,  Rio de Janeiro, RJ, Brazil.}
\date{November 25, 2011}

\begin{abstract}
We present a supersymmetric formulation of  Markov processes, represented by a family  
of Langevin equations with multiplicative white-noise. 
The hidden symmetry encodes equilibrium properties such as fluctuation-dissipation relations. 
The formulation does not depend on the particular prescription to define the Wiener integral.
In this way, different equilibrium distributions,  reached at long times for each prescription, can be formally 
treated on the same footing.
\end{abstract}






 
 
 


\maketitle

\section{LANGEVIN REPRESENTATION OF A
MULTIPLICATIVE WHITE-NOISE PROCESS}

Several years ago it was recognized that some stochastic processes have a hidden 
symmetry~\cite{ParisiSourlas1979,ParisiSourlas1982,Tsvelik1982} called supersymmetry (SUSY). From a mathematical 
point of view, SUSY shows up as an invariance under linear transformations that mix commutative as well 
as auxiliary anticommutative variables in a path integral representation of the process. Physically, 
it encodes equilibrium properties of the system. Some of  the constraints it imposes on correlation 
functions (Ward-Takahashi identities) are related to fluctuation-dissipation theorems~\cite{Chaturvedi1984}. 
This property has acquired a renewed interest due to the growing importance of  stochastic out-of-equilibrium 
systems~\cite{Corberi2007}. In this sense, it is possible to understand out-of-equilibrium dynamics as a 
symmetry-breaking mechanism. SUSY properties have been studied extensively  for additive stochastic 
processes~\cite{Zinn-Justin,Bouchaud1996} and, more recently, for non-Markovian multiplicative processes~\cite{AronLeticia2010}. 
However, the important case of  Markovian multiplicative white-noise 
systems has remained elusive~\cite{arenas2010}. The main difficulty in the SUSY formulation of a Markov process resides in the
great variety of prescriptions to define the Wiener integral, which produces several stochastic evolutions 
with different final steady states. Moreover, time reversal transformations mix different prescriptions.  
In this letter we present a supersymmetric (prescription independent) formulation of this type of process. We show that, 
at the core of SUSY is the non-trivial definition of backward stochastic processes and the 
equilibrium distribution reached at long times. 

Multiplicative white-noise stochastic processes have a wide 
spectrum of applications, not only in physical and chemical  systems~\cite{vanKampen,gardiner}, 
but also in biology~\cite{freund} and even in evolution of economic variables~\cite{Mantegna}. 
The paradigm of multiplicative stochastic dynamics is represented by a Langevin equation with state-dependent 
noise, given in its simplest form by
\begin{equation}
\frac{dx(t)}{dt} = f(x(t)) + g(x(t))\zeta(t).
\label{eq.Langevin} 
\end{equation}
$x(t)$ is a single stochastic variable and $\zeta(t)$ is a Gaussian white noise,
 $\left\langle \zeta(t)\right\rangle   = 0$, $\left\langle \zeta(t)\zeta(t')\right\rangle = \delta(t-t')$.
The drift force $f(x)$ and the diffusion function $g(x)$ are,  in principle,  arbitrary smooth functions of $x(t)$. 
It is well known that the complete definition of the above stochastic differential equation is achieved only upon 
fixing an interpretation for the product $g(x(t))\zeta(t)$ or, more technically, upon the proper definition 
of the Wiener integral~\cite{gardiner}. There are several prescriptions used to define this product, which can be 
summarized in the so called ``generalized Stratonovich prescription''~\cite{Hanggi1978} or 
``$\alpha$-prescription''~\cite{Janssen-RG}.
This definition contains the usual It\^o  ($\alpha=0$) and  Stratonovich ($\alpha=1/2$) prescriptions  as particular cases.
The interpretation of eq.~(\ref{eq.Langevin}) depends on the physics behind a particular application. 
Once the interpretation is fixed, the stochastic dynamics is unambiguously defined. 

The Fokker-Plank equation (FPE) associated with eq.(\ref{eq.Langevin}) can be written as a continuity equation 
$\partial_t P(x,t)+\partial_x J(x,t)=0$, supplemented with the initial condition $P(x,0)=P_{\rm in}(x)$, where 
the probability current is 
$
J(x,t)=[f(x)-(1-\alpha) g(x)\partial_x g(x)] P(x,t)-(g^2(x)/2)\partial_x P(x,t)
$.
We assume that the system converges, at long times, to an {\em equilibrium probability distribution}, 
$P_{\rm eq}(x)=N \exp(-U_{\rm eq}(x))$, computed as a stationary solution of the FPE with zero stationary 
current probability $J_{\rm st}(x)=\lim_{t\to\infty}J(x,t)=0$. The unique solution is 
$U_{\rm eq}(x)=-2 \int^x (f(x')/g^2(x')) dx'+ (1-\alpha) \ln g^2(x) $.
When considering  $f$ as a conservative force, obtained from a potential $V(x)$ as $f=-(1/2) g^2 \partial_x V$, the 
equilibrium distribution has the simpler form, 
\begin{equation}
U_{\rm eq}(x)= V(x)+(1-\alpha)\ln g^2(x).
\label{eq.U}
\end{equation} 
Therefore, the equilibrium distribution depends, not only on the independent functions $f(x)$ and $g(x)$, 
but also on the prescription used to define the stochastic process. 

Note that, only for $\alpha=1$ (H\"anggi-Klimontovich 
pres\-crip\-tion~\cite{Hanggi1982,Klimontovich}), we obtain the usual thermodynamic equilibrium distribution $U_{\rm eq}(x)=V(x)$. 
Provided we know that, for a given physical system,  the equilibrium distribution is of the Boltzmann type with a {\em previously known potential}  $V(x)$, it is possible to determine the stochastic process that converges to this distribution. To do this, the function $f(x)$ is modified by adding a ``spurious drift term'',  $f\to f - (1-\alpha) g \partial_x g$~\cite{Lubensky2007}. In this way, any correlation function is uniquely determined and equilibrium properties do not depend on the value of $\alpha$ (see, for instance, refs.  \cite{Volpe2010,commentVolpe2010,replyVolpe2010} for a discussion on the experimental determination of $f(x)$ and the spurious drift terms in nanoscopic systems).

In this letter, we will  adopt an alternative  
point of view.  We consider a stochastic process modelled by eq.(\ref{eq.Langevin}), with a particular value 
of $0<\alpha<1$.  In this context, the value of $\alpha$ is part of the model and the equilibrium distribution reached at long times is given by eq.(\ref{eq.U}). In this way, we can model stochastic dynamics whose final equilibrium distribution is not necessarily dictated by thermodynamics, such as econophysics applications~\cite{Mantegna}.

\section{TIME REVERSAL AND EQUILIBRIUM PROPERTIES}

A key concept in our approach is the proper definition of time reversal evolution. 
The $\alpha$-prescription is implicitly associated with a particular direction in time. For instance, the It\^o interpretation, also called the  ``prepoint discretization rule'', states that in a discretized time formalism, the Wiener integral $\int g(x(t))\xi(t)dt$ should be computed by evaluating $g(x(t_j))$, using the value of $x(t)$ at the ``initial'' time of each  interval $[t_j,t_{j+1}]$. On the other hand, in the H\"anggi-Klimontovich prescription, or ``postpoint discretization rule'',  we evaluate $g(x(t_{j+1}))$ taking the value of $x(t)$ at the ``final'' time of each  interval. Of course, in a backward evolution the role of ``initial'' and ``final'' are interchanged. In fact, the post-point prescription is sometimes called ``interpretation in the backward sense''~\cite{Lindner2007}. In the $\alpha$-prescription,  the dissipation function is evaluated taking a weighted average of the ``initial'' and ``final'' values of $x(t)$,   $g[x(\tau_j)]=g[(1-\alpha)x(t_j)+\alpha x(t_{j+1})]$.  When considering a backward evolution ($t_j \leftrightarrow t_{j+1}$),  $g[x(\tau_j)]$ is obtained by replacing  $\alpha$ by $1-\alpha$. Thus, if the forward stochastic process is performed in a definite prescription $\alpha$, the backward  trajectories evolve with the dual prescription $1-\alpha$. 
Taking into account that  the equilibrium distribution $U_{\rm eq}$ (given by eq.(\ref{eq.U})) must be the same whenever 
it is reached in a forward ($t\to +\infty$) or backward ($t\to -\infty$) process, the correct definition of the time reversal 
transformation is, 
\begin{equation}
{\cal T} = \left\{
\begin{array}{lcl} 
x(t) &\to & x(-t)   \\ 
\alpha &\to & (1-\alpha) \\
f &\to&  f -\left(1-2\alpha\right)\; g \partial_x g
\end{array}
\right.
\label{eq.T}
\end{equation}
Since $1-2\alpha$ is odd under the transformation $\alpha\to 1-\alpha$, 
${\cal T}^2=I$, as it should be. In the Stratonovich prescription ($\alpha=1/2$), ${\cal T} f= f$ and ${\cal T}$ simply corresponds to a change in the sign of the velocity. In  any other prescription, the definition of the time reversal operator is more involved. 

It is interesting to see how the time reversal transformation is implemented in the path integral formalism. We assign to 
each stochastic trajectory $x(t)$,  with endpoints $(x_i,t_i)$ and $(x_f,t_f)$, a weight 
${\cal P}(x(t)|x_it_i,x_ft_f)=  {\det}^{-1}(g)\;  e^{-S[x]}$.  The ``action'' $S[x]$ is given by~\cite{arenas2010}
\begin{equation}
 S = \int_{t_i}^{t_f} dt \left\lbrace\frac{1}{2g^2} \left[ \frac{dx}{dt} - f + \alpha g \partial_x g\right]^2+\alpha \partial_x f \right\rbrace .
\label{eq.Sfw}
\end{equation}
The conditional probability of the system being in state $x_f$ at time $t_f$, provided  it was initially 
in state $x_i$ at time $t_i$, is given by $P(x_f, t_f|x_i, t_i)=\int {\cal D}x\; {\cal P}(x(t)|x_it_i,x_ft_f)$ with the boundary conditions $x(t_i)=x_i$ and $x(t_f)=x_f$.

Using the transformation (\ref{eq.T}), we compute the time reversed conditional probability  $\hat P(x_i, t_f|x_f, t_i)={\cal T} P(x_f, t_f|x_i, t_i)$, which can be written in the path integral formalism as
$
\hat P(x_i, t_f|x_f, t_i)=
\int {\cal D}x\; {\det}^{-1}(g)\;  e^{-\hat S[x]}
$, 
with the boundary conditions $x(t_i)=x_f$ and $x(t_f)=x_i$  and the time reversed ``action'' $\hat S[x]={\cal T} S[x]$.
Notice that, in general, ${\cal T} P(x_f, t_f|x_i, t_i)\neq P(x_i, t_f|x_f, t_i)$, since the backward stochastic process $(\hat f,g,1-\alpha)$ is not the same, but the dual of the forward one $(f,g,\alpha)$.

After some tedious but direct manipulations, it is possible to write the forward action $S$ (eq.(\ref{eq.Sfw})) as 
$
S= \frac{1}{2}\left.\left(V+\alpha\;\ln g^2\right)\right|_{t_i}^{t_f}+ \tilde S
$, in such a way that $\tilde S$ is time reversal invariant, $\tilde S= {\cal T} \tilde S$.
The time reversed action takes the form  
$
\hat S={\cal T} S= -\frac{1}{2}\left.\left(V+(2-3\alpha)\;\ln g^2\right)\right|_{t_i}^{t_f}+ \tilde S
$.
Thus, the variation of the action under a time reversal transformation  is just a total derivative term (for arbitrary values of $\alpha$),  
$S-\hat S= U_{\rm eq}(x_f)-U_{\rm eq}(x_i)$,
where $U_{\rm eq}(x)$ is the equilibrium potential of eq.(\ref{eq.U})~\footnote{We stress that, for getting this result, it is necessary 
to use the generalized chain rule 
$
dY(x(t))/dt=\partial_x Y (dx/dt) +((1-2\alpha)/2) \partial^2_x Y g^2
$, 
appropriated for a stochastic It\^o calculus with $\alpha$-prescription}.
This property immediately implies the detailed balance relation,
\begin{equation}
P_{\rm eq}(x_i)\; P(x_f, t_f|x_i, t_i) = \hat P(x_f, t_i|x_i, t_f)\;P_{\rm eq}(x_f),
\label{eq.detailedbalance}
\end{equation}
where $P_{\rm eq}(x)=N exp \ (-U_{\rm eq}(x))$. Eq. (\ref{eq.detailedbalance}) is valid for any prescription $\alpha$ and it reduces to the usual detailed balanced relation for additive noise when $\alpha=1/2$ since, in this case,  $\hat P(x_f, t_i|x_i, t_f)=P(x_f, t_i|x_i, t_f) $. 

In the context of stochastic thermodynamics~\cite{seifert2008}, we can compute the increase of entropy in the medium associated with 
one specific trajectory as~\cite{crooks1999,seifert2005},  
$\Delta s_m[x(t)]=\ln\left[{\cal P}(x(t)|x_it_i,x_ft_f)/{\hat {\cal P}(x(t)|x_it_f,x_ft_i)}\right]$.
For each stochastic trajectory beginning and ending in states sampled with the distribution $p(x)$, the total entropy 
production is 
$\Delta s= \ln p(x_i)-\ln p(x_f)- U_{\rm eq}(x_f)+U_{\rm eq}(x_i)$.
In the absence of an explicit time-dependent driving force, the stochastic entropy is a state function which depends only on the initial 
and final states.  Moreover, if we prepare the initial and final states with the equilibrium  distribution $p(x)=P_{\rm eq}(x)$, we 
immediately conclude that $\Delta s=0$, for each stochastic trajectory and for arbitrary values of $\alpha$.  
The relevance of the ``potential'' $U_{\rm eq}$ in  the entropy production was recently recognized in the context of  
the energy representation of a  Brownian system in the Stratonovich prescription~\cite{levkiselev2010}. Here, we show that this concept is much more general, and it does not depend on the specific value of $\alpha$.  

\section{SUPERSYMMETRIC FORMULATION}

N-point correlation functions can be computed by using the generating functional 
\begin{equation}
Z(J)=\int{\cal D}x\; {\det}^{-1}(g) e^{-S[x]+\int_{-\infty}^{\infty}dt' J(t') x(t')}.
\label{eq.ZS}
\end{equation}
$J(t)$ is a source with compact domain, {\em i.\ e.\ }, it adiabatically goes to zero away from  an  interval $(t_i, t_f)$ in 
which we will compute the correlation functions. The action $S$ (eq.(\ref{eq.Sfw})) is computed  over the entire time axis ($-\infty < t < +\infty$). 
To deduce eq.(\ref{eq.ZS}), we have sampled the initial state with the equilibrium distribution and we have assumed an ergodic stochastic evolution, $P_{eq}(x_i)=\lim_{T\to\infty} P(x_i, t_i|x_{-T}, -T)$.  
In this context, total derivative terms in $S[x]$ do not  contribute to the dynamics of any observable.  In this way, a 
system described by equation (\ref{eq.ZS}) is automatically invariant under time reversal, since $S$ and $\hat S={\cal T} S$ differ just in a total time-derivative term. Moreover, if we impose the constraint $x(+\infty)=x(-\infty)$, the action is truly time-reversal-invariant, $S=\hat S=\tilde S$. 

The complex structure of $\tilde S$,  and the fact that, for general $\alpha$, the usual rules of calculus do not apply, 
make calculations rather cumbersome. However, a simplification is possible by extending the space of trajectories, adding one auxiliary ``bosonic'' variable $\varphi(t)$ and a couple of Grassman variables $\bar\xi(t), \xi(t)$. The generating functional (eq.(\ref{eq.ZS})) can be written in the extended functional space as
\begin{equation}
 \mathit{Z}[J] = \int {\cal D}x {\cal D}\varphi {\cal D}\xi {\cal D}\bar{\xi} \ e^{-S[x,\varphi,\xi,\bar\xi] + \int dt J(t)x(t)} \ ,
\label{eq.funcionalgenerador}
\end{equation}
where the action $S$, in the new variables, is given by~\cite{arenas2010},
\begin{eqnarray}
 S&  =&  \int_{-\infty}^{+\infty} \!\!\!\! dt \left\lbrace  -\bar{\xi}(t)\frac{d}{dt}\xi(t) + f'(x)\bar{\xi}(t)\xi(t) 
 +\frac{1}{2}\varphi(t)^2 g(x)^2\right.  \nonumber \\ 
 &+& \left.  i\varphi(t) \left[ \frac{dx}{dt} - f(x) + g(x)\partial_x g(x)\bar{\xi}(t)\xi(t)\right]\right\rbrace .
\label{eq.SGrassman}
\end{eqnarray}
Of course, through functional integration over the auxiliary variables, we obtain the generating functional (\ref{eq.ZS}), with the action given by Eq. (\ref{eq.Sfw}). The $\alpha$ parameter appears in the definition of the equal-time retarded Green's function $\langle\bar\xi(t)\xi(t)\rangle_{\rm R}=\alpha$. It should be noted that, for consistency, the equal-time advanced Green's function should be defined with the dual prescription  $\langle\bar\xi(t)\xi(t)\rangle_{\rm A}=1-\alpha$. In this way, $\langle\bar\xi(t)\xi(t)\rangle_{\rm R}-\langle\bar\xi(t)\xi(t)\rangle_{\rm A}=1-2\alpha$ is odd under time reversal.
The complexity of the stochastic calculus, associated with the definition of the Wiener integral, is now 
codified in the structure of the Grassmann variables. In this way, we can formally work out any calculation 
without explicitly indicating a specific prescription. 

The time-reversal transformation, eq.(\ref{eq.T}), is represented as a linear transformation in the extended  space,
\begin{equation}
{\cal T} = \left\{
\begin{array}{lcl}
x(t) &\to  &  x(-t)  \\
\varphi(t) & \to  & \varphi(-t) -\frac{2i}{g^2}\; \dot x(-t) \\
\xi(t) &\to & \bar\eta(-t)   \\
\bar\xi(t) &\to &- \eta(-t)
\end{array}
\right.
\label{eq.TG}
\end{equation}
where $\eta, \bar\eta$ are the time reversal Grassman variables.  
There is a subtlety in this transformation which involves  boundary conditions. 
While $\lim_{t\to \pm\infty} \bar\xi(t)\xi(t)=\alpha$, the time reversed variables satisfy $\lim_{t\to \pm\infty} \bar\eta(t)\eta(t)=1-\alpha$. 

Computing now the variation of the action under a time-reversal transformation, we obtain 
$
S-\hat S= U_{\rm eq}(x_f)-U_{\rm eq}(x_i)+ \Delta s
$,
where the entropy production associated with each trajectory (in the extended space) is
\begin{equation}
\Delta s=\int_{-\infty}^{\infty} \ln g^2\; \frac{d~}{dt}\left(\bar\xi\xi\right) \; dt\;.
\label{eq.Entropy}
\end{equation}
Interestingly, we observe that the entropy production vanishes due to {\em fermionic number conservation}, or more precisely, due 
to the invariance of the action under global phase transformation of the Grassman variables.

The path-integral formalism is useful to make evident (otherwise hidden) symmetries of the stochastic process. For instance, 
the action given by eq.(\ref{eq.SGrassman}) is invariant under the transformation, 
$\delta x = \bar\lambda \xi$,  $\delta \xi=0$, 
$\delta \bar\xi = i\bar\lambda \varphi$, $\delta \varphi=0$,  
where $\bar\lambda$ is an anticommuting parameter.  This nilpotent transformation  ($\delta^2=0$) is the famous BRS~\cite{bechi1976} 
symmetry, discovered in the context of quantization of gauge theories. In the present context, it simply enforces  probability 
conservation,  $Z(0)=1$. There is another set of important symmetries related with equilibrium properties that, together with BRS, is called supersymmetry. To display it explicitly, it is convenient to introduce a ``natural'' response variable 
$\tilde\varphi= g^2\left(\varphi+i \partial_x\ln g^2\bar\xi\xi\right)$ and to re-scale the Grassman variables 
$\eta=g \xi$, $\bar\eta=g \bar\xi$, in such a way that the  transformation 
$(x,\varphi,\xi,\bar\xi)\to (x,\tilde\varphi,\eta,\bar\eta)$, has a trivial Jacobian, ${\cal D}\varphi {\cal D}\xi {\cal D}\bar{\xi}={\cal D}\tilde\varphi {\cal D}\eta {\cal D}\bar{\eta}$. 
In these variables, it is simpler to formally compute response functions. To see this,  we slightly perturb the system out of equilibrium  $V\to V+h(t) x$, and compute the linear response
\begin{equation}
 R(t,t')=\left.\frac{\delta \langle x(t)\rangle_h}{\delta h(t')}\right|_{h=0}=\langle x(t) i \tilde\varphi(t')\rangle,
\end{equation}
where the last correlation function should be computed with the action $S(x,\tilde\varphi,\eta,\bar\eta)$.

It is convenient to collect the transformed dynamical variables  in the definition of a simple 
scalar superfield, 	
\begin{equation}
\Phi(t,\theta,\bar\theta) = x(t)+ \bar \theta\eta(t)+ \bar\eta(t)\theta  +i \tilde\varphi(t) \bar\theta \theta, 
\end{equation}
where we have introduced two ``temporal''  Grassman variables $\theta$ and $\bar\theta$. 
The system can be described in terms of two superpotentials  
$V(\Phi)$ and $\Gamma(\Phi)$, where the ``diffusion potential'' is introduced as  $g^{-1}(x)=\partial_x\Gamma(x)$.
Then, the time-reversal-invariant action $\tilde S$, can be re-written in terms of $\Phi(t,\theta,\bar\theta)$ as
\begin{equation}
\tilde S=\int dtd\theta d\bar\theta\; \left\{\bar D\Gamma[\Phi] D\Gamma[\Phi]+\frac{1}{2} V[\Phi]\right\},
\label{eq.SSUSY}
\end{equation}
where we have defined the covariant derivatives 
$\bar D=\partial_\theta$, $D = (1/2)\partial_{\bar\theta}-\theta \partial_t$, 
satisfying  $D^2=\bar D^2=0$ and $\left\{D,\bar D\right\}=-\partial_t$.

The action of eq.(\ref{eq.SSUSY}) (and, of course, that one of eq.(\ref{eq.SGrassman}))  has a SUSY whose generators are
\begin{equation}
 Q= \partial_{\bar\theta} \mbox{~~,~~} \bar Q= (1/2)\partial_\theta+\bar\theta\partial_t
\mbox{~~,~~} \left\{Q,\bar Q \right\}=\partial_t \ .
\end{equation}
The graded algebra 
$\left\{Q, D \right\}=\left\{Q, \bar D \right\}=\left\{\bar Q, D \right\}=\left\{\bar Q, \bar D \right\}=
Q^2= \bar Q^2=0$, 
guarantees that the variation of the action is a total derivative.

Each of the three generators of SUSY $\{ Q, \bar Q, \partial_t\}$ imposes several non-perturbative constraints on correlation functions. $\partial_t$ induces time translations and means that any two-point correlation function depends on time differences.  
On the other hand, $Q$, the generator of $\bar\theta$ translations, represents the usual BRS symmetry, responsible for the probability conservation. Moreover, the invariance of the action generated by $\bar Q$ imposes, for instance, 
\begin{eqnarray}
\lefteqn{
\langle\Phi( t_1,\theta_1,\bar\theta_1)\Phi( t_2,\theta_2,\bar\theta_2)\rangle=} \nonumber  \\
&&\langle\Phi(t_1+\epsilon \bar\theta_1,\theta_1+\epsilon/2,\bar\theta_1)\Phi(t_2+\epsilon \bar\theta_2,\theta_2+\epsilon/2,\bar\theta_2)\rangle
\label{eq.PhiPhi}
\end{eqnarray} 
where $\epsilon$ is an arbitrary Grassmann parameter. Eq.(\ref{eq.PhiPhi}) implies, when written in components, 
the fluctuation dissipation relation  $R(t,t')\sim \partial_t \langle x(t)x(t')\rangle\theta(t-t')$.

For additive processes, the diffusive potential is linear, $\Gamma(x)\sim x$, and Eq.(\ref{eq.SSUSY}) 
reduces to the usual action defined with a single superpotential $V(\Phi)$. In this case, the tadpole theorem~\cite{arenas2010} 
guarantees that the stochastic evolution does not depend on $\alpha$. However, multiplicative processes (non-linear $\Gamma$), 
induce derivative couplings in the superfield. These couplings are responsible for the  $\alpha$-dependent evolution that leads 
to the equilibrium distribution of Eq.(\ref{eq.U}). 

\section{SUMMARY AND CONCLUSIONS}

Summarizing, we have shown a hidden SUSY in a general  multiplicative white-noise stochastic process. SUSY does not depend 
on the particular definition of the Wiener integral ($\alpha$) and encodes equilibrium properties, such as  
fluctuation-dissipation relations. The main ingredient in the SUSY formulation is the proper definition of time reversal 
and the $\alpha$-dependent equilibrium distribution. Finally, we presented a covariant formulation based on two superpotentials 
related with the drift force and with the diffusion function. 
The SUSY structure is particularly useful to study multiplicative white-noise dynamics, since any specific computation can be worked out in a unified way, without the specification of a particular value of $\alpha$ (nor a particular rule of calculus). 
Moreover, the   nonperturbative constraints  imposed by SUSY help to select the relevant diagrams that should be taken into account in a perturbative calculation.

\acknowledgments
The Brazilian agencies CNPq and FAPERJ are acknowledged for partial  financial support.
Z.G.A. was partially supported by the Latin American Center of Physics, CLAF, under the collaborative program CLAF-ICTP.



%

\end{document}